\documentclass [preprint,aps,showpacs]{revtex4}
\usepackage[final]{graphics}
\usepackage{amssymb}
\usepackage{amsfonts}
\usepackage{epsfig}
\usepackage{graphicx}

\begin{document}
\title{Quasi-Stationary Distributions for Models of Heterogeneous
   Catalysis}
\author{Marcelo M. de Oliveira\footnote{mancebo@fisica.ufmg.br} and Ronald
Dickman\footnote{dickman@fisica.ufmg.br}}
\address{Departamento de F\'{\i}sica, Instituto de Ci\^encias Exatas,\\
Universidade Federal de Minas Gerais \\
C. P. 702, 30123-970, Belo Horizonte, MG - Brazil }

\date{\today}

\begin{abstract}

We construct the {\em quasi-stationary} (QS) distribution for two
models of heterogeneous catalysis having two absorbing states:
the ZGB model for the oxidation of CO, and a version
with noninstantaneous reactions. Using a mean-field-like
approximation, we study the quasi-stationary surface coverages,
moment ratios
and the lifetime of the QS state. We also derive an improved,
consistent one-site mean-field theory for the ZGB model.

\end{abstract}

\maketitle

\section{Introduction}

The Ziff-Gullari-Barshard(ZGB) model \cite{ziff86} for
the oxidation of carbon monoxide (CO) on a catalytic surface
exhibits phase transitions between an active steady state and absorbing or
``poisoned'' states, in which the surface is saturated either by
oxygen (O) or by CO. This model and its variants have stimulated
much interest; many studies of the ZGB model have been
published using deterministic mean-field equations
\cite{dickman86,fish-titu89,dumont90} and Monte Carlo simulations.
For reviews see \cite{evans91,zhdanov94,marr-dick99}.)
In the present work we analyze quasi-stationary properties of the
model.  In the process, we revisit the one-site mean-field theory
and formulate an improved version thereof.

The surface reactions described by the ZGB and allied models follow the
Langmuir-Hinshelwood mechanism \cite{engel-ertl79},
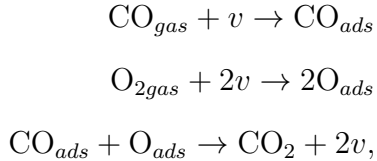
\begin{eqnarray}
\mbox{CO}_{gas}+v\to \mbox{CO}_{ads} \nonumber \\
\mbox{O}_{2 gas}+2v \to 2\mbox{O}_{ads} \nonumber \\
\mbox{CO}_{ads}+\mbox{O}_{ads}\to \mbox{CO}_{2}+ 2v, \nonumber
\end{eqnarray}

\noindent where $v$ denotes a vacant lattice site, and the
subscripts indicate the state (gaseous or
adsorbed) of the molecule. The product CO$_2$ is understood to desorb
immediately on formation. CO$_{gas}$ molecules arrive at a rate
$Y$ per site while O$_2$ arrives at rate $(1-Y) \equiv \overline{Y}$.

Monte Carlo simulations \cite{ziff86} reveal that the model
exhibits two phase transitions.  For $Y<y_1=0.391$ the lattice
eventually poisons with O while for $Y>y_2=0.5256$
it poisons with CO. For $y_1<Y<y_2$ the system attains a
reactive steady state. The first transition is found to be
continuous while the second is strongly discontinuous.
Mean-field-like descriptions based on cluster approximations \cite{dickman86}
reproduce qualitatively the behavior of the system in
the thermodynamic limit.

Large but finite systems appear to be stationary on
any reasonable time scale (away from the phase boundaries),
but eventually end up in one of the
absorbing states (O-poisoned or CO-poisoned), via a fluctuation
with a very small, but nonzero probability.
Mean-field theories ignore such fluctuations, and so are incapable
of treating finite systems. Since simulations and other numerical
methods usually study finite systems, it is
useful to develop approximate theoretical descriptions that
account for finite system size.
Experimental study of nanoscale catalytic systems is also
of current interest \cite{suchorski01}. The notion of a quasi-stationary
(QS) distribution has proved to be a powerful tool in this
context. For such models, the QS distribution describes the
asymptotic (long-time) properties of a finite system conditioned
on survival \cite{yaglom47,ferrari,nasell01}.
The quasi-stationary properties converge to the stationary
properties when the system size $N \to \infty$.

In Ref. \cite{dickman-vidigal02}
QS distributions (QSD) are derived for various continuous-time
Markov processes, e.g., the contact process and the
Malthus-Verhulst process. In Ref. \cite{atman02} the QSD
for a discrete-time process is derived, using the
Domany-Kinzel cellular automaton as an example. In all of these
studies, mean-field-like approximations are used to find the QSD.
Previous studies have focused on continuous phase transitions to
an absorbing state, with the exception of Schl\"{o}gl's second
model \cite{schlogol72}, known to exhibit a discontinuous
transition in mean field approximation, which was shown in Ref.
\cite{dickman-vidigal02} to exhibit a bimodal QSD at the transition.
Such a bimodal
distribution (e.g., of the order parameter) is a hallmark of a
discontinuous phase transition. The phase transition in
Schl\"{o}gl's second model, with nearest-neighbor interactions,
is however known to be {\it continuous} in $d<4$ dimensions
\cite{grassberger82}. Hence, aside from the particular interest of
the ZGB model in the context of catalysis, the present work
presents a QS analysis of a model known to exhibit a discontinuous
transition.

The balance of this paper is organized as follows. In Sec.~II, we
briefly review the mean-field theory for the ZGB model at the site
level, introducing modifications that take into account the
prohibition against O-CO nearest-neighbor (NN) pairs on the surface.
In Sec.~III, we
define the QS probability distribution and show how to derive it
in an efficient way. This method is used to generate the
results reported in Sec.~IV; results for the case of noninstantaneous
reactions (NI model) are
also presented. Sec.~V contains a summary of
our results. The Appendix treats the ZGB model on
a complete graph.

\section{Site Mean-Field Approximations}

The simplest mean-field description of the ZGB model is the site
approximation. By assuming spatial homogeneity and neglecting all
correlations between the occupancies of different sites, one
obtains a closed description in terms of the CO and O coverages,
i.e., the fraction of sites
occupied by CO(O). To simplify the notation, from this point we
will denote a CO molecule by $c$, a vacant site by $v$, and an O atom
by $o$. The coverages will be denoted by $\theta_{i}$, for $i=c,$
$v$ or $o$, with the normalization $\theta_c+\theta_o+\theta_v=1$.

In the ZGB model reactions between
adsorbed $c$ and $o$ are instantaneous, so that there is a strict
prohibition against $c$-$o$ NN pairs.
Previous formulations of the site
approximation \cite{dickman86,fish-titu89,dumont90} effectively
ignored this prohibition, leading to the appearance of an
absorbing state in which a small fraction (about 0.2\%) of the sites are occupied
by $o$, while the rest bear $c$ molecules.  Elimination of this
nonphysical effect is one motivation for reformulating the
approximation; providing a systematic basis for the QS analysis
is another.  In the absence of any prohibitions, the site approximation
treats each site as statistically independent, so that the joint
probability $\theta_{ij}$ for a NN pair of sites to be in states
$i$ and $j$ is $\theta_i \theta_j$.  Our problem is to formulate
an approximation that treats sites as independent, {\it except}
for the fact that $\theta_{oc} = \theta_{co} = 0$.
(We use $\theta_{ij}$ to denote the joint probability that, on the
square lattice, the leftmost or lower site of the pair is in state
$i$ and the other site in state $j$.
By symmetry $\theta_{ji} = \theta_{ij}$.)

On a square lattice of $N$ sites (with periodic boundaries, so that
each site possesses exactly four NNs), there are $2N$ NN pairs.
Let $N_i$ be the number of sites in state $i$ and $N_{ij}$ the number
of NN pairs with states $i$ and $j$, using the same convention as
for $\theta_{ij}$.  Then we have the relations

\begin{equation}
N_v = \frac{1}{2} (N_{vv} + N_{vo} + N_{vc})
\label{Nv}
\end{equation}

\begin{equation}
N_o = \frac{1}{2} (N_{oo} + N_{vo})
\label{No}
\end{equation}
and
\begin{equation}
N_c = \frac{1}{2} (N_{cc} + N_{vc})
\label{Nc}
\end{equation}
Letting $\theta_{o|o}$ denote the conditional probability that a site
bears an $o$ atom, given that a NN of this site does, we have
\begin{equation}
N_{oo} = 2 N_o \theta_{o|o} = 2 N_o \frac{\theta_o}{\theta_o + \theta_v}
\label{Noo}
\end{equation}
where we used the fact that if a site bears $o$, then its NN cannot
harbor $c$.  By the same reasoning
\begin{equation}
N_{ov} = 2 N_0 \frac{\theta_v}{\theta_o + \theta_v}
\label{Nov}
\end{equation}

\begin{equation}
N_{cc} = 2 N_c \frac{\theta_c}{\theta_c + \theta_v}
\label{Ncc}
\end{equation}
and
\begin{equation}
N_{cv} = 2 N_c \frac{\theta_v}{\theta_c + \theta_v}
\label{Ncv}
\end{equation}

Combining Eqs. (\ref{Nv}), (\ref{Nov}) and (\ref{Ncv}) we then have
\begin{equation}
N_{vv} = 2 N_v \left[1  - \frac{\theta_o}{\theta_o + \theta_v}
- \frac{\theta_c}{\theta_c + \theta_v} \right]
\label{Nvv}
\end{equation}
Note that in general $N_{vv} < 2N_v \theta_v $, the value it would
have in the absence of the prohibition against $c$-$o$ NN pairs.
In order to eliminate such pairs, vacancies must, as it were, be
redistributed to create more $v$-$o$ and $v$-$c$ pairs than would
be present under completely random mixing, and in this redistibution
the number of $v$-$v$ pairs is diminshed.

Now using $\theta_{ij} = N_{ij}/2N$ and
$\theta_{i|j} = \theta_{ij}/\theta_j$, we may proceed to assign
rates to the various adsorption/reaction events in the model.
Consider, for example, nonreactive adsorption of $c$.  This
occurs at an intrinsic rate of $Y$, and requires a vacant site
whose four NNs are free of $o$.  The probability that a site
does not harbor $o$, given that its NN is vacant, is
$1 - \theta_{o|v} = \theta_v/(\theta_o + \theta_v)$, so that the
rate $W_1$ (or expected number of events per site and unit time) of
nonreactive $c$ adsorption is
\begin{equation}
W_1 = Y \theta_v \left( \frac{\theta_v}{\theta_o + \theta_v} \right)^4
\equiv Y \theta_v \alpha
\label{W1}
\end{equation}
The rates for the remaining events,
listed in Table I, are found in the same manner
(for further details see \cite{dickman86,ben-kohler92}).
We use the notation
\begin{equation}
\theta_{vv} = \theta_v \left[1  - \frac{\theta_o}{\theta_o + \theta_v}
- \frac{\theta_c}{\theta_c + \theta_v} \right]
\label{tvv}
\end{equation}
and
\begin{equation}
\gamma =  \left( \frac{\theta_v}{\theta_c + \theta_v} \right)^3
\label{gamma}
\end{equation}

From the rates listed in Table I, one readily derives the following equations of
motion for the coverages:
\begin{eqnarray}
\label{1site-zgb} \dot{\theta_o} & = &
2 \overline{Y} \theta_{vv} \gamma
-Y\theta_v (1-\alpha) \nonumber \\
\dot{\theta_{c}} & = &
Y\theta_v \alpha -2 \overline{Y} \theta_{vv} (1-\gamma)
\end{eqnarray}

Numerical integration of these equations reveals that
for $Y < Y_s^>$ = 0.56101, starting from an empty lattice, the system
attains an active steady state. $Y_s$ marks a spinodal point, the
limit of stability of the active phase \cite{marr-dick99}. The
site approximation does not reproduce the O-poisoning transition,
yielding $Y_s^<= 0$.

The present formulation is an improved version of the one-site approach
presented in \cite{dickman86}. In the latter formulation one finds
$Y_s^>=0.5615$, but, as noted,
the final state (for $Y > Y_s^>$) has a small admixture of $o$
in the $c$-poisoned phase,
in contradiction to the prohibition against $c$-$o$ NN pairs.
The present formulation eliminates this flaw and
provides a slightly improved estimate for $Y_s^{>}$.

In the preceding discussion we consider the thermodynamic limit
(there is no reference to the system size). The model
exhibits an active phase below the spinodal value $Y_s^>$. In
finite systems it is not possible to avoid poisoning. But for
large system sizes, we expect, for a range of $Y$, a long-lived
metastable state, which we call the quasi-stationary state.
Increasing the system size, the lifetime of the QS state goes to
infinity, and its properties become equivalent to that of true
active stationary state. In the next section we use the QS
probability distribution in order to study finite systems.

\section{Quasi-stationary probability distributions}

The dynamics of a stochastic process such as the ZGB model is governed by
a master equation (ME) for the probabilities
in state space; the latter consists
of all possible configurations of the system. Since the transition rates
of the model do not satisfy
detailed balance, one cannot determine the
steady-state probability distribution without solving the ME. We
may write a simplified ME, based on a cluster approximation, which,
depending on the truncation level, includes
certain fluctuations.

In the simplest case, one truncates the ME at the site level. This
means that the state of the system is specified by the site
occupation numbers $c$, $o$ and $v$, without regard to the
relative positions of the adsorbed species on the surface.
(Evidently, the fluctuations preserved at this level are
those in the total occupancy.)
Consider a square lattice of $N$ sites, so
that $c+o+v=N$. The state of the ZGB model in the site approximation is given by
the pair $(c,v)$; the model defines a
Markov chain in the $c$-$v$ plane (see Fig. 1), with $\theta_c = c/N$, etc.
The O-poisoned and CO-poisoned
(absorbing) states are $(0,0)$ and $(N,0)$, respectively.

In the site approximation, the probability distribution
$P_{c,v}(t)$ follows the ME

\begin{equation}
\dot P_{c,v}(t) = \sum_{c',v'}W_{(c',v')\to(c,v)}P_{c',v'}(t)-
P_{c,v}(t)\sum_{c',v'}W_{(c,v)\to(c',v')}
\end{equation}
where $W_{(c,v)\to(c',v')}$ is the transition rate.
\noindent In terms of the ZGB rates shown in Table 1, we have:
\begin{equation}
\label{ME1site}
 \dot{P}_{c,v} =
W_1P_{c-1,v+1}+W_2P_{c,v-1}+W_3P_{c,v+2}+W_4P_{c+1,v}+
W_5P_{c+2,v-2}-\sum_{i=1}^{5}W_iP_{c,v}
\end{equation}

\noindent In Eq. (\ref{ME1site}) it is understood that each of the $W_i$ is
to be evaluated for the same values as the associated probability
factor (that is, in the first term, $W_1$ must be evaluated for
$(c-1,v+1)$, etc.).

``Poisoning'' in the ZGB model corresponds to the system falling
into one of the absorbing states (see Fig.1). This means that, for
a finite system, the stationary probability distribution is nonzero
only for absorbing configurations, that is,
\begin{equation}
\label{final}
\lim_{t \to \infty} P_{c,v} (t) = p \delta_{(c,v),(0,0)}
+ (1-p) \delta_{(c,v),(N,0)}
\end{equation}
(The weight $p$ depends on $N$, $Y$, and the initial distribution.)

We introduce the QSD via the hypothesis that, as $t\to \infty$, the probability distribution
{\em conditioned on survival} attains a time-independent form.
Then the QS probability distribution
$Q_{c,v}$ is given by

\begin{equation}
Q_{c,v}=\lim_{t \to \infty}P_{c,v}(t)/S(t)
\end{equation}
where $S(t)$ is the {\em survival probability}, i.e., the probability
that the system has not fallen into the absorbing state up to time $t$.
In the Appendix, we present results on the QS distribution for the
ZGB model on a {\it complete graph}, which admits an exact description
in terms of a single variable.

One method for generating the QSD is by integrating the
master equation until the distribution $P_{c,v}^s \equiv
P_{c,v}(t)/S(t)$, (with $S(t)=1-P_{0,0}(t)-P_{N,0}(t)$) attains a
time-independent form. For moderately large systems ($N \geq 100$),
standard schemes such as the fourth-order Runge-Kutta method
\cite{numerical}, require long integration times, because a
small time step is needed to avoid numerical instabilities.
An alternative method is
available \cite{dickman02}, based on writing the evolution in the
form:

\begin{equation}
\label{it1}
 \dot{P}_{c,v}(t) = - W_{c,v}P_{c,v}(t)+A_{c,v}(t),
\end{equation}
\noindent where
\begin{equation}W_{c,v} \equiv
\sum_{c',v'}{W_{(c,v)\to (c',v')}}
\end{equation}
\noindent and
\begin{equation}
 A_{c,v}(t)=\sum_{v'\geq 1,c'}{W_{(c',v') \to (c,v)}p_{c',v'}(t)}
\end{equation}

Suppose the probability distribution has attained a QS form at
some time $t$. Using normalization, $\sum_{v \geq
1,c}{Q_{c,v}(t)=1}$ and Eq. (\ref{it1}), we have

\begin{equation}
A_{c,v}=(W_{c,v}-A_0)P_{c,v},
\end {equation}

\noindent where
\begin{equation}
A_0 \equiv \sum_{v'>1,c'}{W_{(c',v') \to
(c,0)}}P_{c,v}= -\sum_{v\geq 1,c}\dot P_{c,v}(t)
\end{equation}
is the decay rate of the survival probability.

For the ME, Eq. (\ref{1site-zgb}),
$W_{c,v}=\sum_{i=1}^{5}{W_i}$ and
\begin{equation}
A_{c,v}=W_1P_{c-1,v+1}+W_2P_{c,v-1}+W_3P_{c,v+2}+W_4P_{c+1,v}
+W_5P_{c+2,v-2}
\end{equation}
 The decay rate is $A_0=W_1P_{N-1,1}+W_3P_{0,2}$.
In the quasi-stationary regime we have
\begin{equation}
\label{it2} Q_{c,v}=\frac{A_{c,v}}{W_{c,v}-A_0} \ \  (v \geq 1),
\end{equation}

\noindent which suggests the following
iteration scheme:
\begin{equation}
P'_{c,v}=a P_{c,v}+(1-a)\frac{A_{c,v}}{W_{c,v}-A_0},
\end{equation}

\noindent  where $a$ is a parameter and $A_{c,v}$ is evaluated
 using the distribution $P_{c,v}(t)$
The new distribution $P'_{c,v}$ must be normalized after
each iteration. In this way, one can construct the QS state from
any normalized initial distribution $P_{c,v}(t)$. This
\emph{iterative method} exhibits good convergence for $a=0.5$.

\section{Results}

We construct the QS distributions for the ZGB model at the
one-site level, via both the Runge-Kutta method and the iterative
scheme, which as expected yield
the same QS distribution. Using $a=0.5$, we find that
the iterative method converges to the QS distribution about 100
times faster than via the RK method using the same precision. This
gain in efficiency is associated with the small time step required
to maintain numerical stability in the RK scheme. In Fig. 2, we
show a typical QS distribution, $Q_{c,v}$, for the ZGB model near
the spinodal point. It is clearly bimodal, with one maximum
at $c=N\!-\!1$, near
the CO-poisoned state, and another, broader one that
corresponds the active state.

Figure 3 shows the time evolution of the coverages
(conditioned on survival), for two different initial conditions, and
$Y$ fixed at the spinodal value $Y_s^>$ obtained from the MF theory
of Sec. II. It is clear that, although the relaxation to
the QS values is very different in the two cases, both reach
reach the same QS coverages. The relaxations to the QS state
for $Y< Y_s^>$ and $Y> Y_s^>$ are compared in Fig. 4 (Since
the iterative method provides only the QS distributions, the
evolution is calculated via the Runge-Kutta method in this case.)

It is also possible to discuss QS properties for $Y>Y_s$, where an
active stationary state does not exist, even in the thermodynamic
limit. This is a finite-size effect, analogous to a
nonzero magnetization in the Ising model above the critical
temperature, on a finite lattice. This effect was also
reported for the contact process on a complete graph
\cite{dickman-vidigal02}.

The marginal QS distribution $Q_c$ (Fig. 5) becomes bimodal as $Y\to Y_s^>$,
consistent with a discontinuous phase transition and the
emergence of a hysteresis loop when $N\to \infty$.
The lifetime $\tau$ of the QS state is given by
\begin{equation}
\tau=1/A_0.
\end{equation}
As seen in Fig. 6, $\tau$ becomes very large as we increase $N$,
for $Y < Y_s^>$. The
lifetime increases exponentially with $N$ (Fig. 6 - inset),
as found for the CO poisoning time on finite lattices
\cite{ben-avra90}.

Fig. 7 shows the mean number of adsorbed CO molecules,
$\langle c\rangle $, and of
vacant sites $\langle v\rangle $, in the QS state. It also
displays the QS moment ratios $m_c=\langle
C^2\rangle /\langle C\rangle ^2$ and $m_v=\langle V^2\rangle
/\langle V\rangle ^2$ as functions of $Y$. The quantity $m$ is
analogous to Binder's reduced fourth cumulant \cite{binder81} at
an equilibrium critical point.  At a continuous phase transition
to an absorbing state, we expect $m$ for the order parameter
to attain a universal, size-independent value, due to scaling
of the probability distribution \cite{rdjaff,rdmunoz}.
In the present case of a discontinuous phase transition, $m_c$ and $m_v$
can be expected to grow due to the bimodal nature of the probability
distribution.  Very large values of $m_v$ can be attained as the coverge
$\theta_v$ tends to its minimum, $1/N$.
For the finite systems studied here,
the peaks of $m_c$ and $m_v$ do not fall
at the same values of $Y$, as would be
expected for an infinite system. The positions of the maxima
appear to approach one another as $N \to \infty$;
an extrapolation to $N \to \infty$ confirms the peak values
occur at $Y_s^>$ as given by mean-field theory.

\subsection{Noninstantaneous reactions}

One of many possible extensions for the ZGB model that have been
proposed relaxes the condition of instantaneous reactions between adsorbed O and CO.
\cite{dumont90}. The key difference between this NI model and the ZGB model
is that there is no longer
any prohibition against $c$-$o$ NN pairs.  There is now a finite
mean reaction time for O and CO adsorbed at NN sites to form CO$_2$;
the reciprocal of the reaction time is the reaction rate
$R$.  (We continue to assume instantaneous desorption of CO$_2$.) The
transitions are listed in Table II; rates are obtained as in
Table I.
(The prefactor 4 in rate $W_3$ is
the coordination number of the square lattice.)

In this case, the mean-field equations are:
\begin{eqnarray}
\label{Rfin-motion}
\dot{\theta_o} &=& 2(1-Y){\theta_v}^2-4R \theta_o\theta_c \nonumber \\
\dot{\theta_c} &=& Y\theta_v-4R \theta_o\theta_c
\end{eqnarray}

A simple analysis of equation (\ref{Rfin-motion}) shows that
$Y_s^> \sim \sqrt{2R}$ as $R \to 0$, grows monotonically with $R$,
attaining the limiting value
$Y_s^> = 2/3$ as $R \to \infty $. This value is different from
the spinodal found in the site-approximation for the ZGB model, which
involves instantaneous reactions. Since $\lim_{R\to \infty}Y_s^> \neq
Y_{s,ZGB}^> $, the limit of infinite reaction rate is singular. The
singular nature of this limit derives from the fact that the
instantaneous reaction condition imposes certain constraints on
configurations, thereby altering the algebraic form of the
mean-field equations (a more extensive discussion may be found in
\cite{dumont90}). In particular, for $R \to \infty$, $\theta_o$
and $\theta_c$ cannot both be nonzero, whereas both coverages are
nonzero in the active phase of the ZGB model.

The probability distribution for the NI model follows the master
equation:
\begin{equation}
\dot P_{c,v}(t) =
R_1P_{c-1,v+1}+R_2P_{c,v+1}+R_3\left[P_{c+1,v-1}+P_{c,v-1}\right]
                     -\left[ R_1+R_2+2R_3 \right] P_{c,v}(t).
\end{equation}

We turn now to a discussion of the QS properties, obtained using
the iterative method based on Eq.~(\ref{it2}). The lifetime of the
QS state, in the reactive window, increases with $R$ (see Fig.~8).
The displacement of the maximum position of $\tau$ (as a function
of $Y$) parallels the variation in $Y_s^> (R)$. The maximum
lifetime (i.e., varying $Y$) tends to a finite value, different
from that from the original ZGB model, when $R\to \infty$. This is
because the flux to the absorbing configurations, which determines
the lifetime of the QS state, and which, in the site
approximation, is given by $A_0=W_1 Q_{N-1,1}+W_3 Q_{0,2}$, is
reduced as $R$ increases, since the latter reduces the QS
probabilities $Q_{N-1,1}$ and $Q_{0,2}$.

Figs. 9 and 10 show the QS coverages $\theta_c$ and $\theta_v$.
As $R$ increases the transition becomes progressively ``harder,"
i.e., the coverage $\theta_c$ just below the transition approaches
zero while that just above approaches unity.
In the limit $R\to \infty$, the $c$ coverage is a step-function:
$\theta_c = \Theta(Y-Y_s^>)$.
As before, the QS coverages
tend to the macroscopic curves (furnished by mean-field theory) when
$N \to \infty$. Due to the low $c$ coverage in the active phase for
large $R$, the moment ratios $m_c$ differ strongly from those
of the ZGB model. But as always, the peak in $m_c$ tends
to the spinodal as $N\to \infty$ (see Fig. 11).

\section{Summary}

We study the quasi-stationary properties of the ZGB model
for the catalytic oxydation of CO,
and of a related model with noninstantaneous reactions. The QS
distribution is constructed numerically via a bivariate extension
of the iterative method \cite{dickman02}.  For the simple case of
a complete graph, we obtain exact (numerical) results for QS
properties.
Although our analysis of the models with NN reactions on a square lattice is
based on site-approximations, which cannot capture critical
fluctuations, the approach allows one to put some fluctuations
``back into'' mean-field theory,
specifically, fluctuations in the overall coverages, since we
assume spatial homogeneity. This enables us
to investigate finite-size effects and moment ratios that are out
of reach with other approaches.  We derive a revised site approximation for
the ZGB model that respects the prohibition against $c$-$o$ NN pairs,
eliminating the inconsistency noted in previous formulations
\cite{dickman86,fish-titu89,dumont90}, while providing a reasonably good
estimate of the spinodal point.

The present study confirms two fundamental properties of the ZGB and
NI models: (1) the discontinuous phase transition is signaled by a
bimodal QS probability distribution; (2) the lifetime of the QS state
grows exponentially with system size, in the reactive phase.  The latter
implies that the system will typically spend a long time in the
QS state, following a relatively brief transient, except near the
spinodal, where relaxation slows down.  We have also noted sharply
increased values of the moment ratios $m_c$ and $m_v$ in the vicinity
of the transition.  Since the CO$_2$ production rate is closely related to
$\theta_v$, one should expect to observe large fluctuations in product
concentrations in the gas phase, as the spinodal is approached.  One may
envision using these fluctuations as part of a control scheme that adjusts
the CO partial pressure (represented by the parameter $Y$ in the models),
to optimize production while avoiding poisoning.

An interesting result is that at the level of mean-field theory, the ZGB model
is not equivalent to the infinite-reaction-rate limit of the model with
noninstantaneous reactions.  The ZGB model is characterized by the
constraint that any $c$-$o$ NN pair react before any further adsorption events,
a condition that is also realized in the NI model, for sufficiently large $R$.
Mean-field theory is however incapable of representing this situation, since,
in the absence of a prohibition against $c$-$o$ pairs, the NI model in the $R \to \infty$
limit simply prohibits simultaneous nonzero coverages of O and CO.  Since truly
instantaneous reactions do not exist, one might suppose that the NI model,
with an appropriately large rate $R$, provides the correct description of fast
(but not instantaneous) surface reactions.  In mean-field approximation however,
the large-$R$ limit of the NI model appears {\it less realistic} than the ZGB model,
which permits simultaneous presence of adsorbed O and CO.  In this sense, the
choice of model in the description of cooperative kinetics should involve the
approximation that one intends to use, not only the intrinsic nature of the
model.

QS analysis promisses to be a useful technique in the study of catalysis
models on nanometer scales \cite{suchorski01}.  Possible directions for
future work
include the study of metastability via QS distributions, and Monte
Carlo simulations of the QS state, in catalytic models and other
processes exhibiting absorbing states.
\vspace{1em}

{\small{\bf ACKNOWLEDGMENT}}
\vspace{1em}

This work was supported by CNPq and CAPES, Brazil.

{\small {\bf APPENDIX: ZGB MODEL ON A COMPLETE GRAPH}}
\vspace{1em}

A {\it complete graph} is one in which all sites are neighbors.  When formulated
on such a structure, a stochastic lattice model defines a stochastic process
defined by one or a few variables, thereby permitting an exact analysis,
as in the case of the contact process \cite{dickman-vidigal02}.
Since each site interacts with all others, the behavior is mean-field-like.
Due to the prohibition against $c-o$ NN pairs, in the ZGB model
on a complete graph, only one species
($c$ or $o$) may be present at any moment. Letting $n=c-o$, we can describe a
system of $N$ sites completely by a single variable, with
$n = -N$ representing the O-poisoned state and $n=N$ the CO-poisoned state.
The number of vacant sites is $v = N-|n|$.

Each adsorption event leads either to $n \to n+1$ or to
$n \to n-2$.  The transition rates are
\begin{equation}
W_{n\to n+1}=Y\left(1-\frac{|n|}{N}\right) \nonumber \\
\end{equation}
and
\begin{equation}
W_{n\to
n-2}=\overline{Y} \left(1-\frac{|n|}{N}\right)\left(1-\frac{|n|+1}{N}\right).
\end{equation}

The process obeys the master equation:
\begin{eqnarray}
\dot{P}_n(t)&=&Yu_{n-1}P_{n-1}(t)+\overline{Y}u_{n+2}\left(u_{n+2}-\frac{1}{N}\right)P_{n+2}(t)\nonumber
\\
 & &-u_n\left[Y+\overline{Y}\left(u_{n}-\frac{1}{N}\right)\right]P_n(t),
\end{eqnarray}
\noindent where $u_n=\left(1-\frac{|n|}{N}\right)$.
The macroscopic equation for this process \cite{ben-avra90} is
given by:
\begin{equation}
\label{macro}
\dot{\phi}(t)=Y(1-|\phi (t)|)-2\overline{Y} (1-|\phi (t)|)^2.
\end{equation}
where $\lim_{N\to\infty}{\phi(t)=n(t)/N}$. In this limit,
 the system poisons with CO when $Y \geq 2/3$. For $Y<2/3$ there is a
nontrivial fixed point at $\phi ^*=1-(1-Y)/2Y$ witch is the
attractor if $\phi(0)>-\phi *$. The point $\phi ^*$ is the reactive
stationary state for an infinite system.

Analyzing the master equation via the iterative method described
in Sec. III, we obtain the QSD shown in Fig. 12. The QSD changes
qualitatively as $Y$ approaches the spinodal. For small $Y$, the
distribution has a single maximum near $\phi ^*$. For $Y \simeq
2/3$ the distribution becomes bimodal, with one peak near $\phi
^*$, related to the active phase, and another at $n=N-1$, related
to the CO-poisoned state. Fig. 13 shows the QS activity for this
model; as $N\to \infty$, it tends to the macroscopic limit (dashed
line). As verified for the site approximation, the lifetime grows
exponentially with $N$ in the active region ($0<Y<2/3$) and
linearly in the nonreactive region.

\newpage

\noindent{\bf TABLES}
\vspace{1em}

\begin{table}
\caption{Site mean-field theory for the ZGB model}
\begin{tabular}{l l l}
\hline \hline
Event    & Rate  & $(c,v)$ to\\
\hline
 $v+c\downarrow \to c$   & $W_1 = Y \theta_v \alpha $   & $(c+1,v-1)$\\

 $v+c\downarrow \to v$   & $W_2= Y \theta_v (1-\alpha)$ & $(c,v+1)$\\

 $vv+oo\downarrow \to oo \;\;\;\;\;\;$ & $W_3 =
 \overline{Y} \theta_{vv} \gamma^2 $  & $(c,v-2)$ \\

 $vv+oo\downarrow \to vo$    & $W_4 =
 2 \overline{Y} \theta_{vv} \gamma  (1-\gamma) \;\;\;\;\; $ & $(c-1,v)$ \\

 $vv+oo\downarrow \to vv$   & $W_5= \overline{Y}
 \theta_{vv} (1-\gamma)^2$  & $(c-2,v+2)$\\

\hline \hline
\end{tabular}
\end{table}

\begin{table}
\caption{Site mean-field theory for the finite reaction-rate
model}
\begin{tabular}{c l l }
\hline \hline
Event & Rate & $(c,v)$ to  \\
\hline
 \ $c\downarrow   \;\;$    & $W_1$ = $\theta_v Y$ & $(c+1,v-1)$   \\
 \ $oo\downarrow  \;\;$    & $W_2$ = $2\theta_v^2\overline{Y}$ & $(c,v-2)$   \\
 \ $oc\uparrow    \;\;$    & $W_3$ = $4R\theta_c\theta_o \;\;$ & $(c-1,v+2)$   \\
\hline \hline
\end{tabular}
\end{table}

\newpage

\noindent{\bf FIGURE CAPTIONS}
\vspace{1em}

\noindent FIG. 1.  Transitions in the ZGB model (site
approximation). Circles denote absorbing states.
\vspace{1em}

\noindent FIG. 2.  A typical bimodal QS distribution
$Q_s(c,v)$ near the spinodal point ($Y=0.56101$).
\vspace{1em}

\noindent FIG. 3. Time evolution of the mean values {\em
conditioned on survival} for $Y=0.56101$ and system size $N=64$. Initial conditions, from top to bottom:
half $c$ and half $v$; full $v$; half $o$ and half $v$. Dashed lines show the
respective unconditioned values.
\vspace{1em}

\noindent FIG.4.Relaxation for the QS mean values of $c$ for
$Y=0.50$, $Y=0.56101$ and $Y=0.60$, from bottom to top. $N=64$.
Dashed lines: unconditioned mean values
\vspace{1em}

\noindent FIG. 5. Marginal distribution $Q_c$ for Y=.54, .55, .56,
.57, .58.
\vspace{1em}

\noindent FIG. 6. Lifetime $\tau$ of the QS state as
function of CO arrival rate $Y$, for the ZGB model.
System sizes $N=36,64,100,256,400,512$ from bottom to top.
Inset: Lifetime as function of the system size $N$ for $Y=0.50$.

\vspace{1em}

\noindent FIG. 7. Mean values {\em conditioned on survival}
$\langle c \rangle$, $\langle v\rangle$,
 and respective moment ratios, $m_c$ and $m_v$ for the
original ZGB model without desorption. System size
$N=16,36,64,100,256,400,700$.

\vspace{1em}

\noindent FIG. 8. Lifetime $\tau$ of the QS state for the
NI model. System sizes $N=100$ (lower
solid lines); $N=256$ (dashed lines); $N=400$ (upper solid lines).
Reaction rates $R=0.1, 1, 10, and 10^6$ from left to right. The
dashed line is the lifetime for the
ZGB model. Inset: detail of the same graph with $N=100$,
showing the lifetime for the ZGB model (dashed line).

\vspace{1em}

\noindent FIG. 9. QS coverage $\theta_c$
for the NI model; reaction rates as in Fig. 8.
System sizes $N=100$ (solid lines) and $N=400$ (dashed lines).

\vspace{1em}

\noindent FIG. 10. QS vacancy fraction $\theta_v$ for the NI model, as in Fig. 9.

\vspace{1em}

\noindent FIG.11. Moment ratio $m_c$ for NI
model.  Reaction rates and system
sizes as per Fig. 9.

\vspace{1em}

\noindent FIG. 12. QSD for the ZGB model on a complete graph
of $100$ sites.

\vspace{1em}

\noindent FIG. 13. Quasistationary activity for the ZGB on a
complete graph. The dashed line represents the macroscopic
limit, Eq. (\ref{macro}).

\vspace{1em}

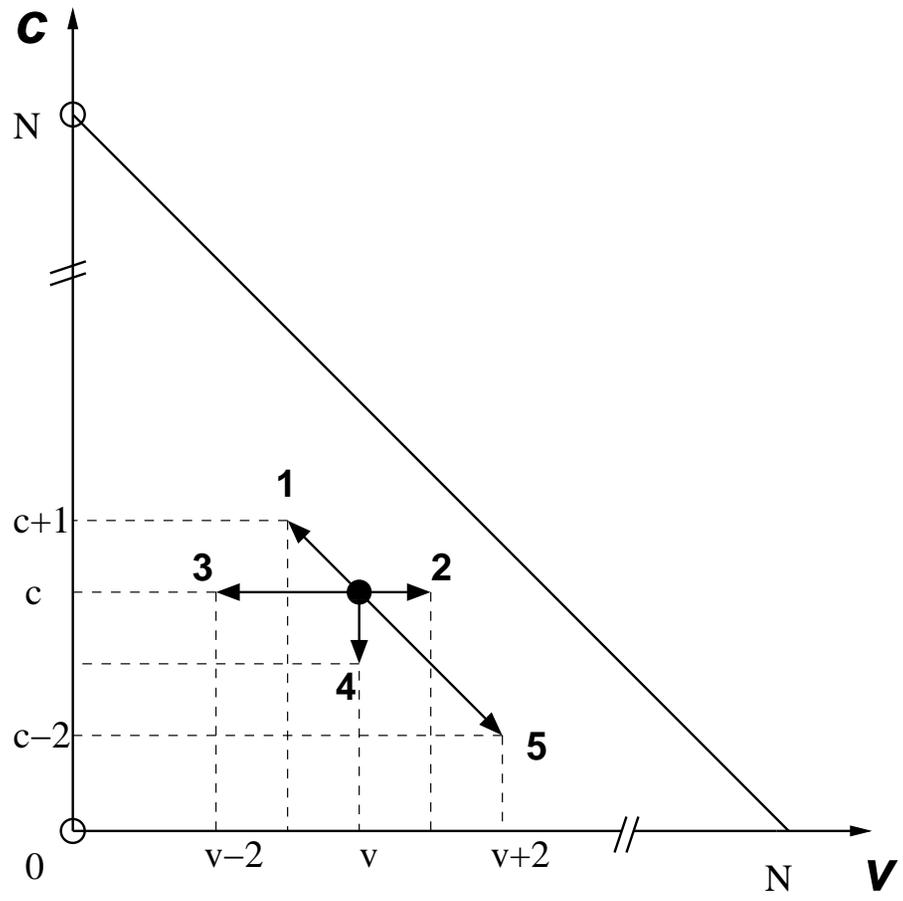
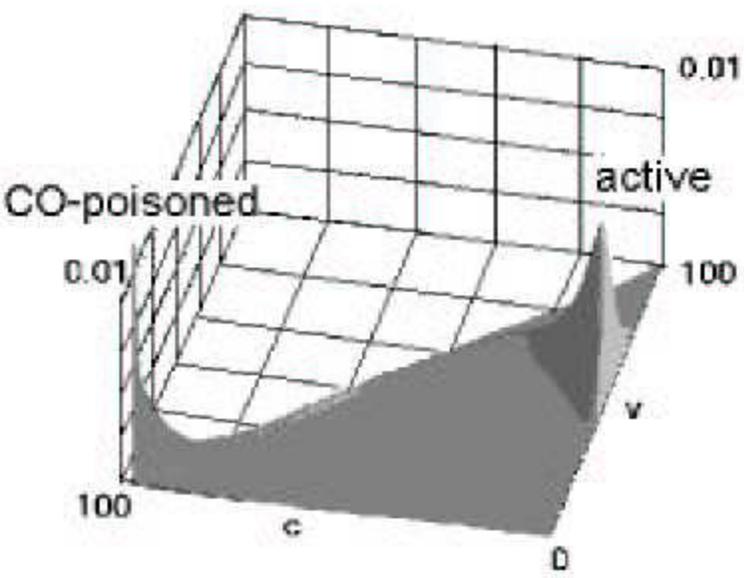
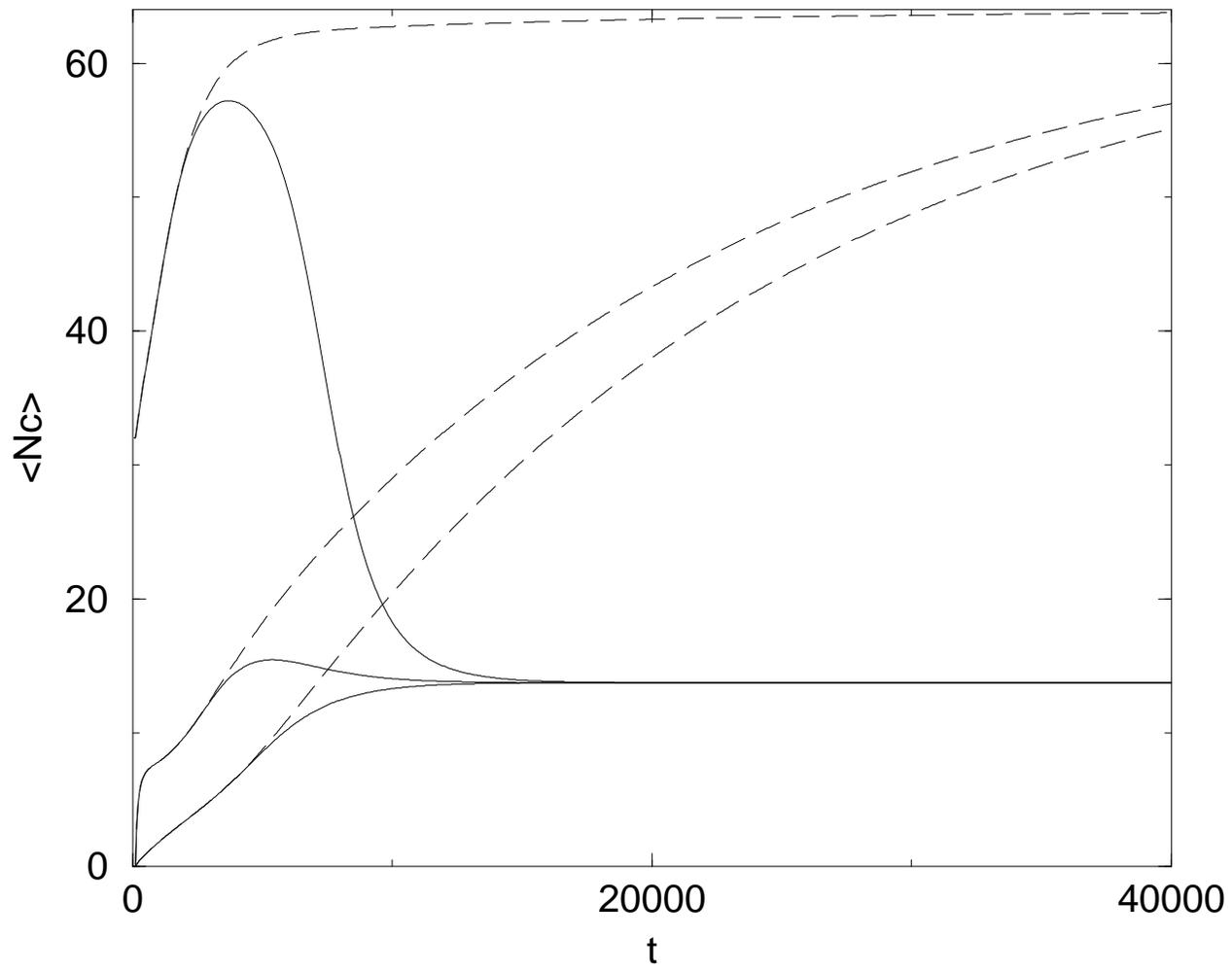
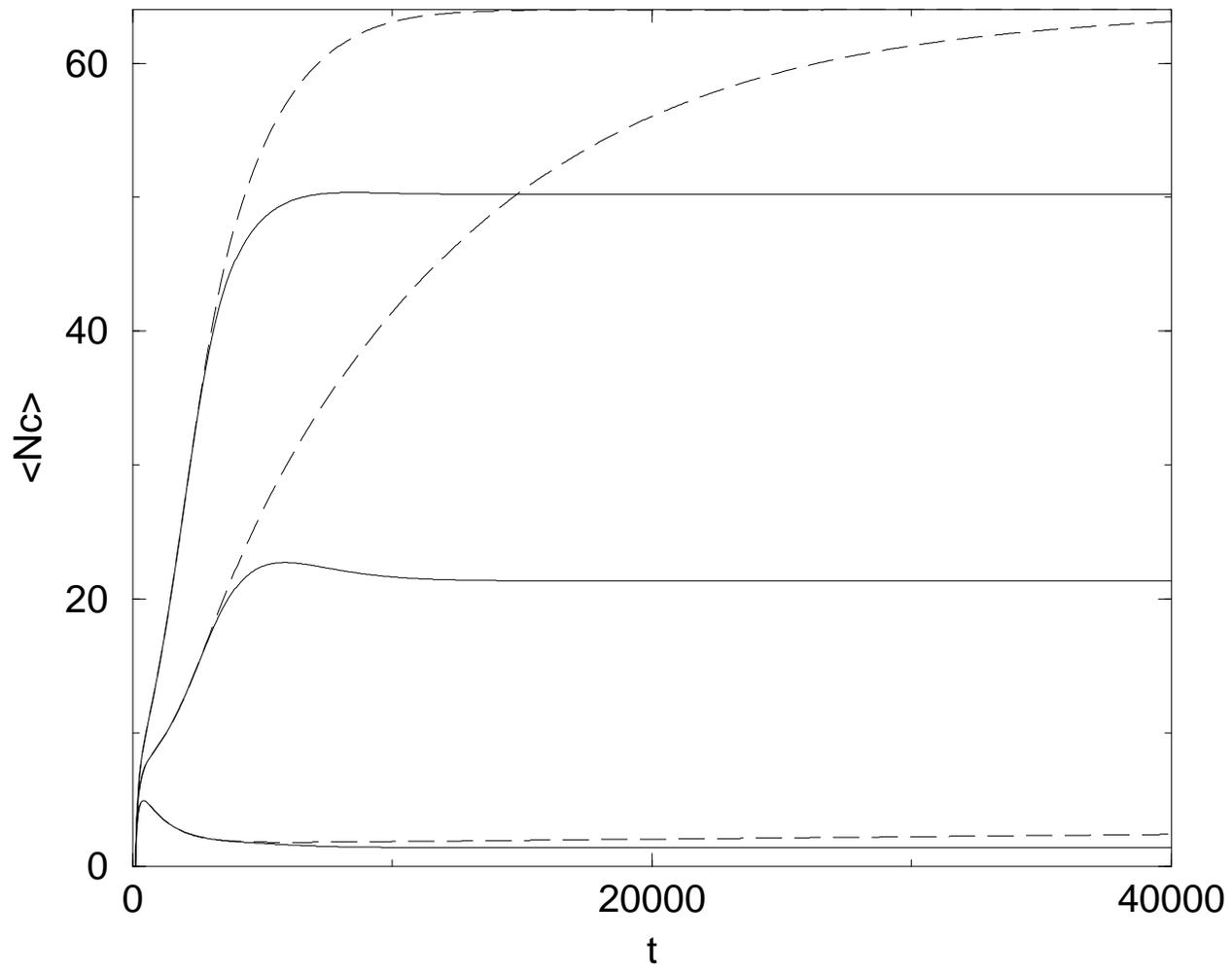
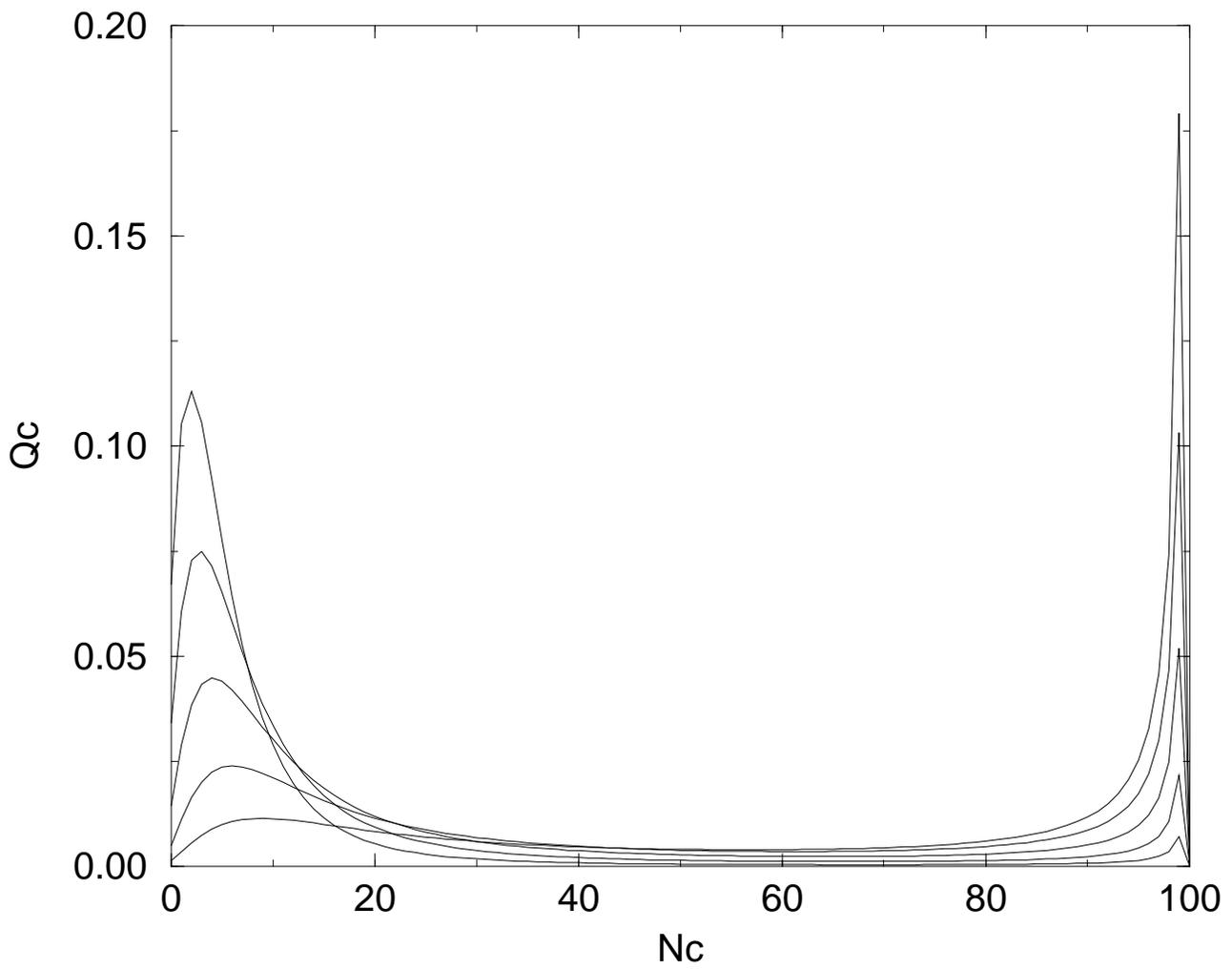
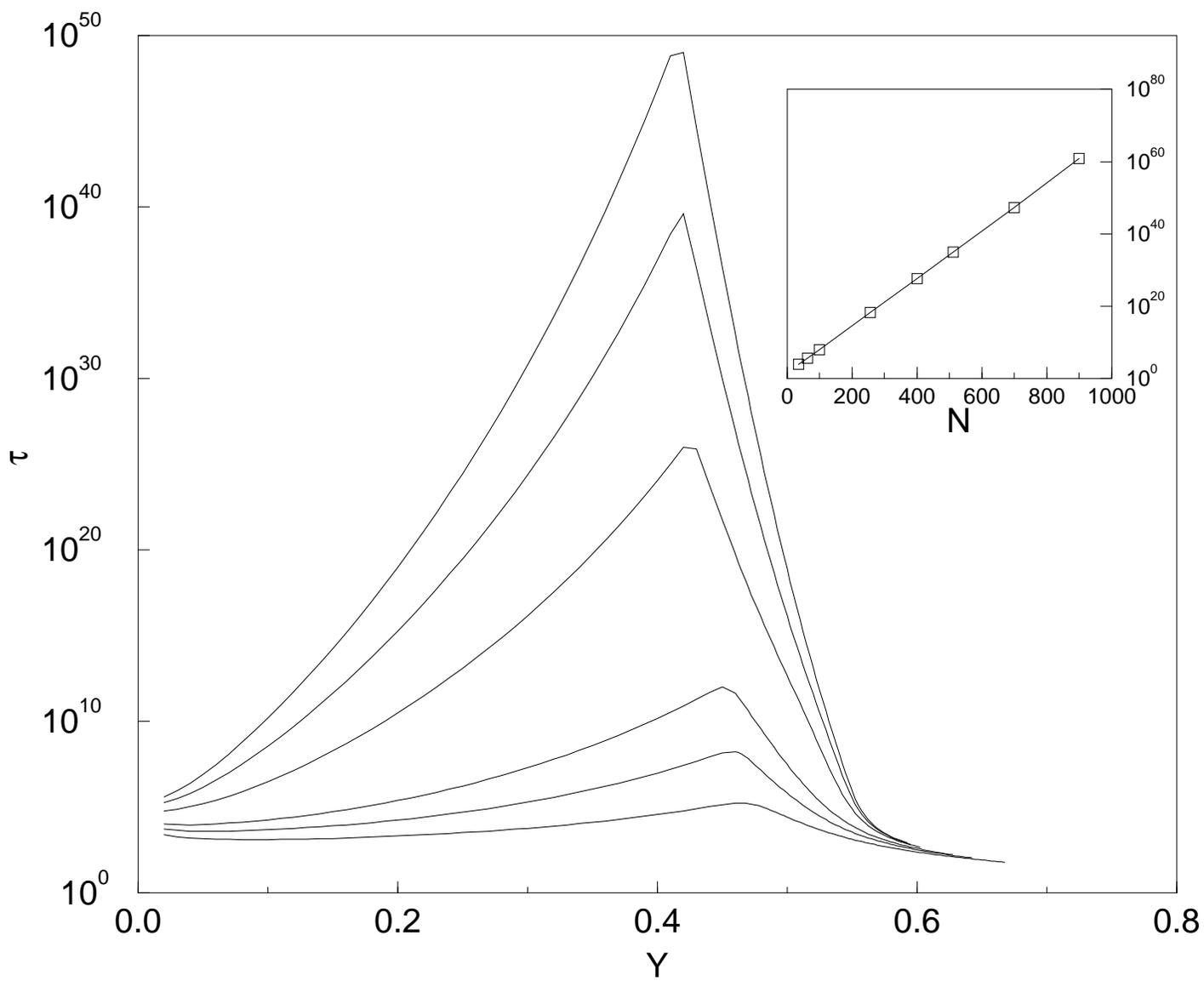
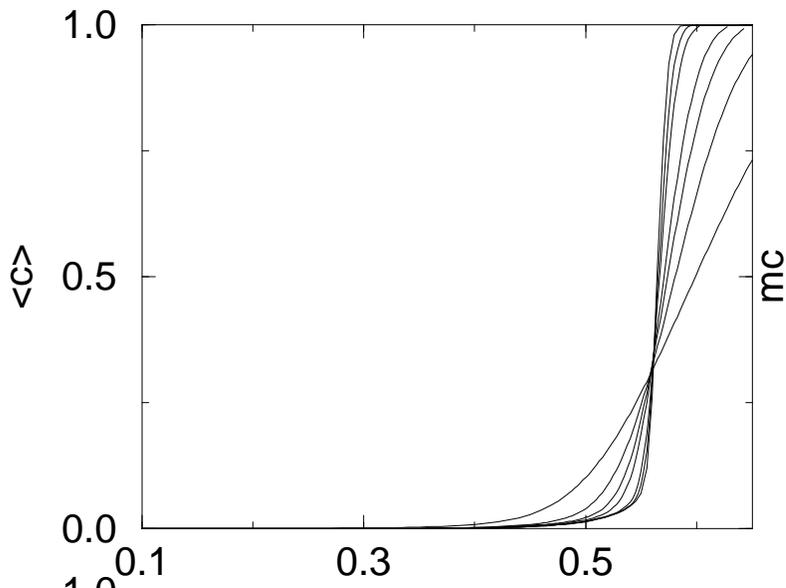
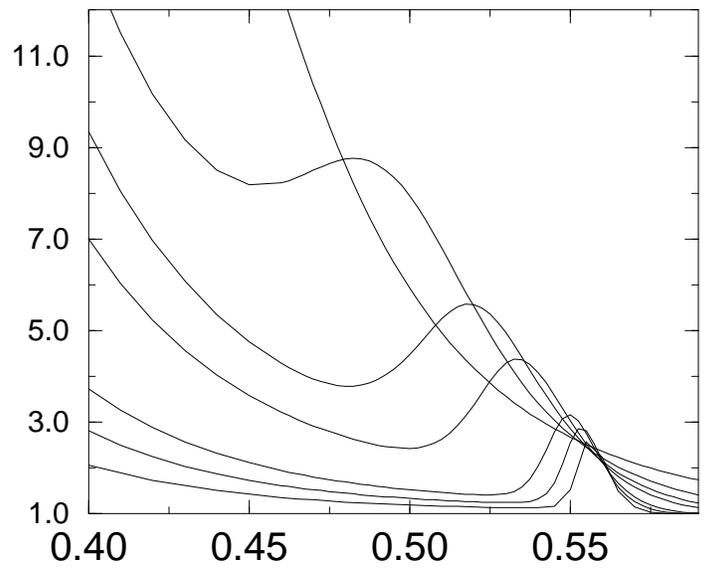
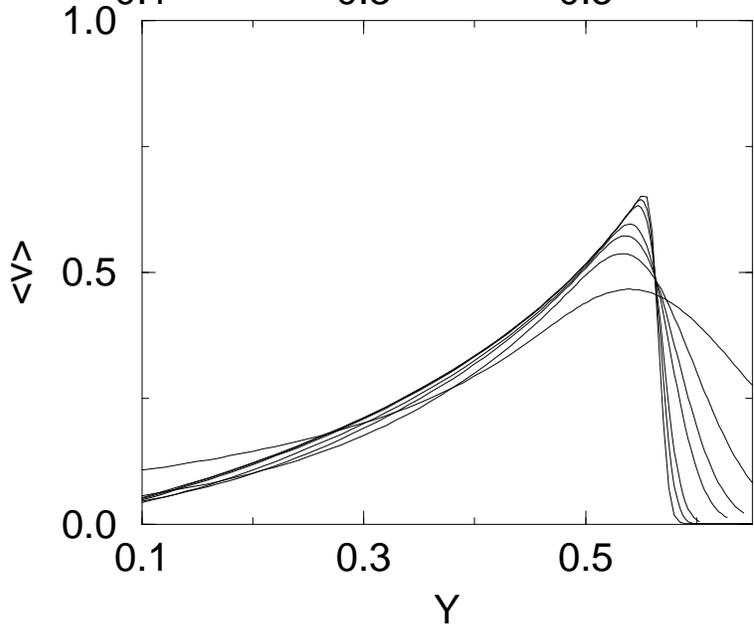
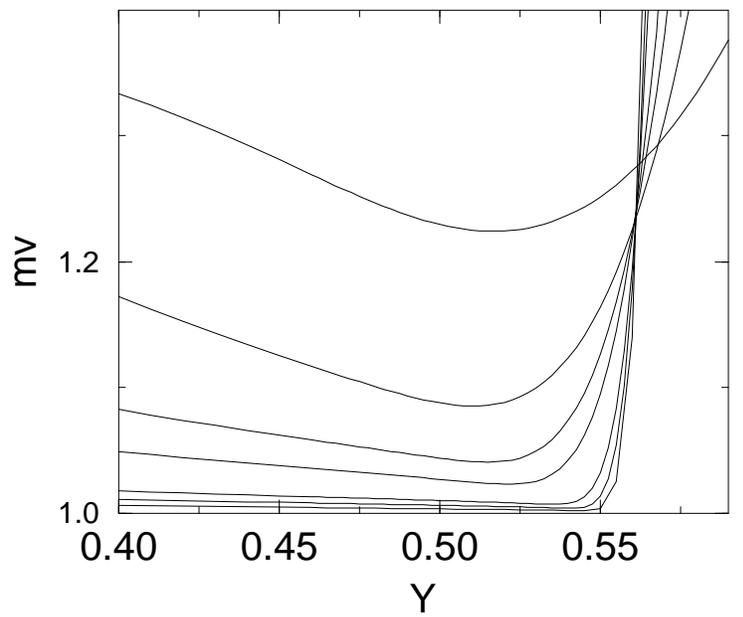
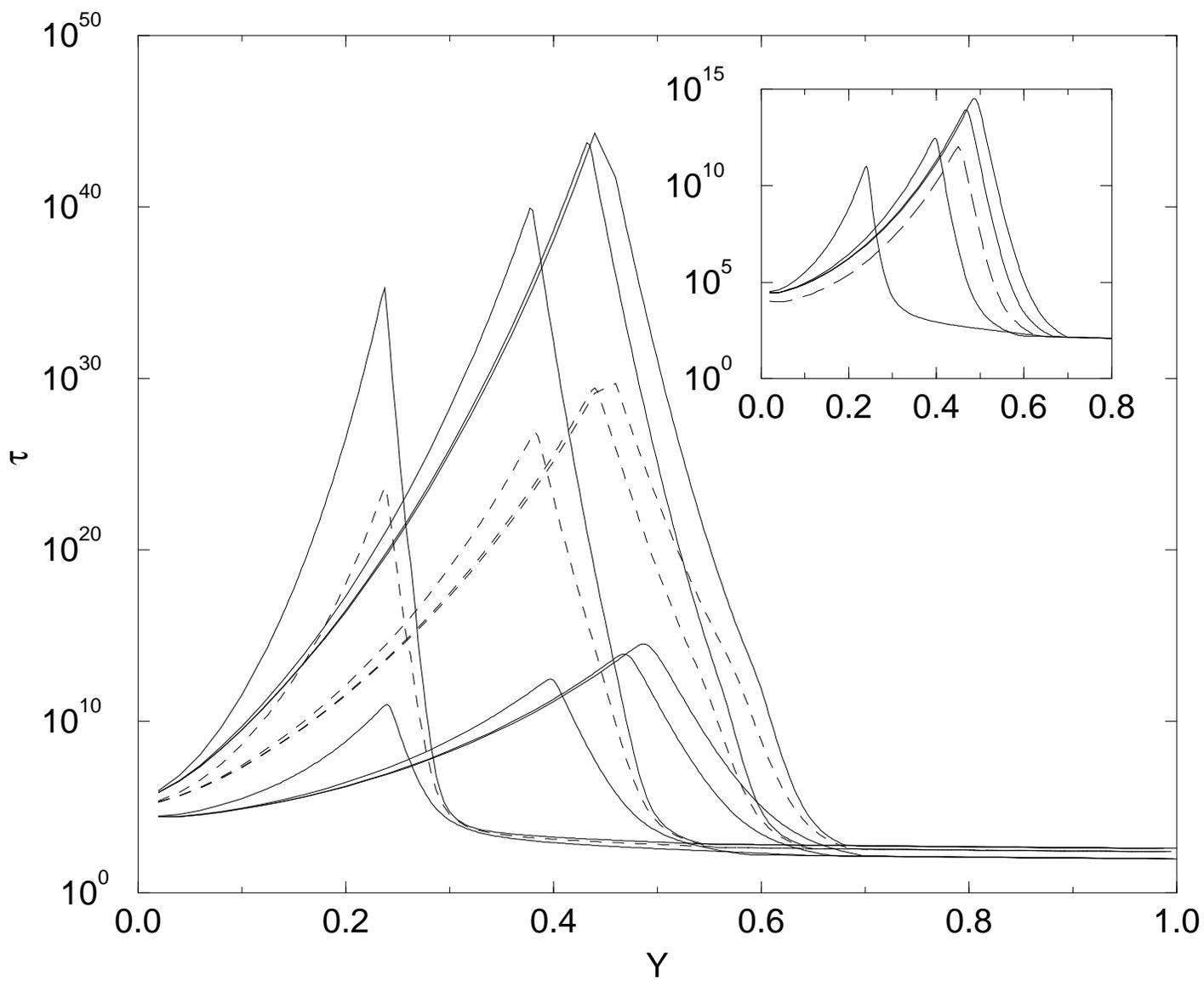
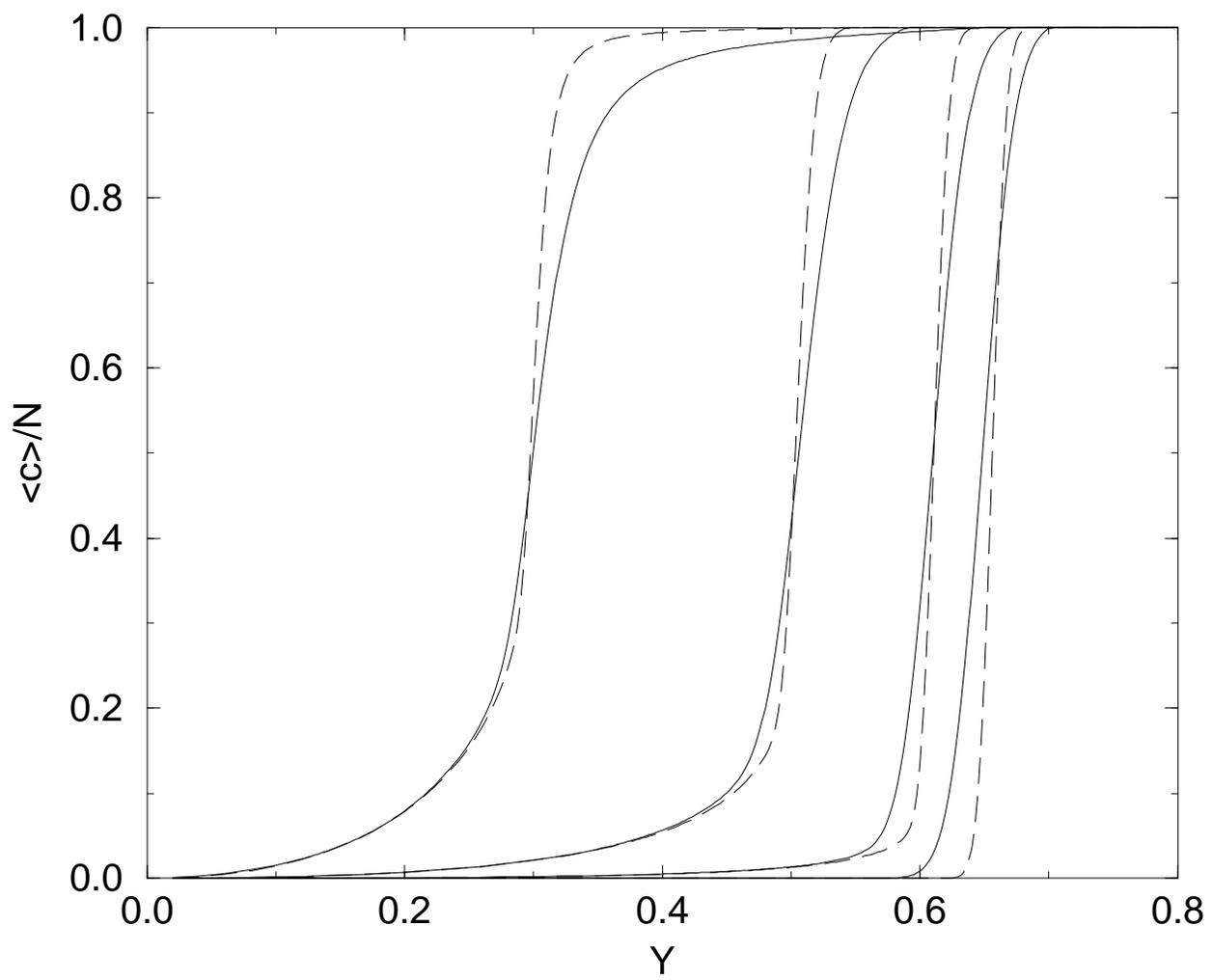
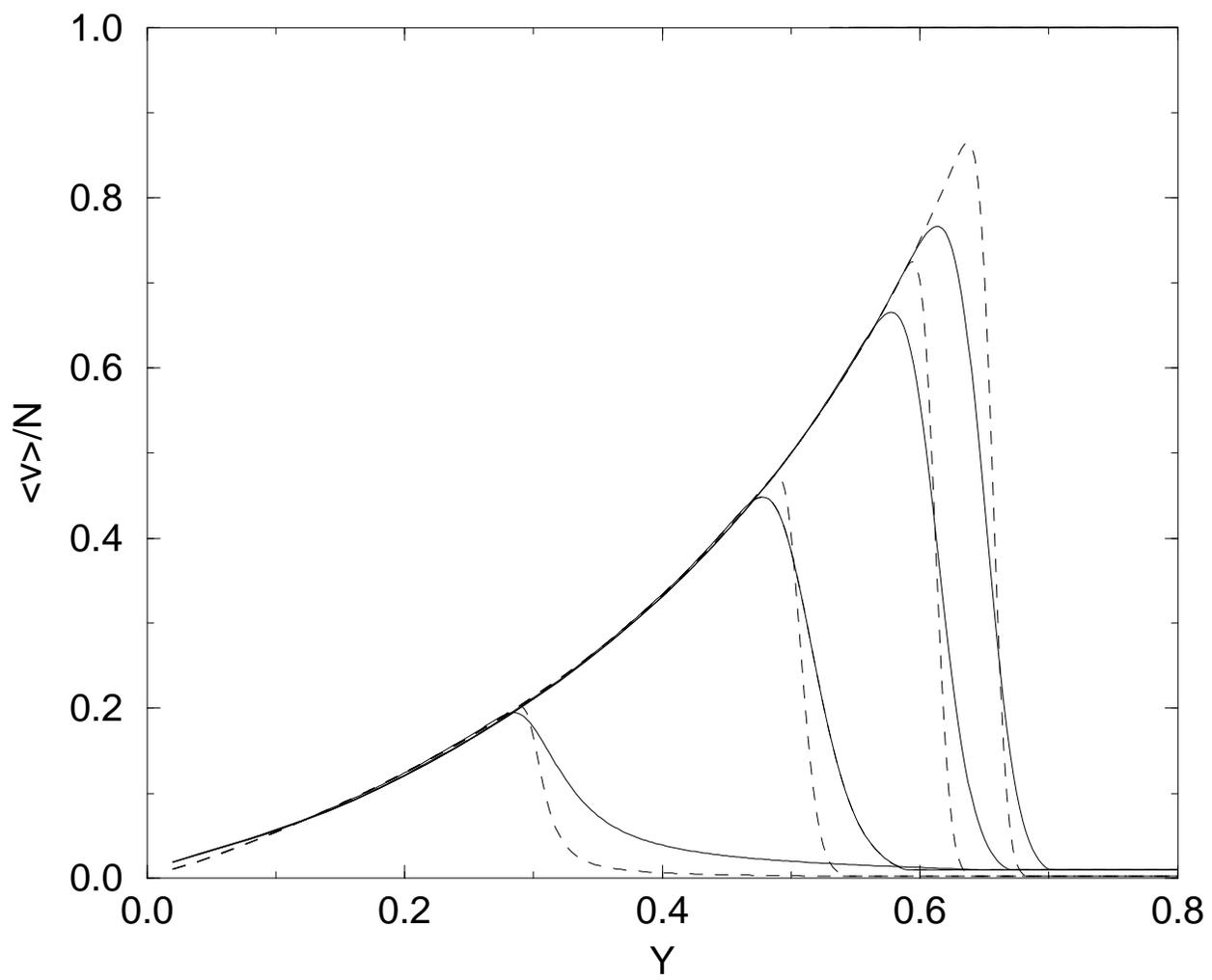
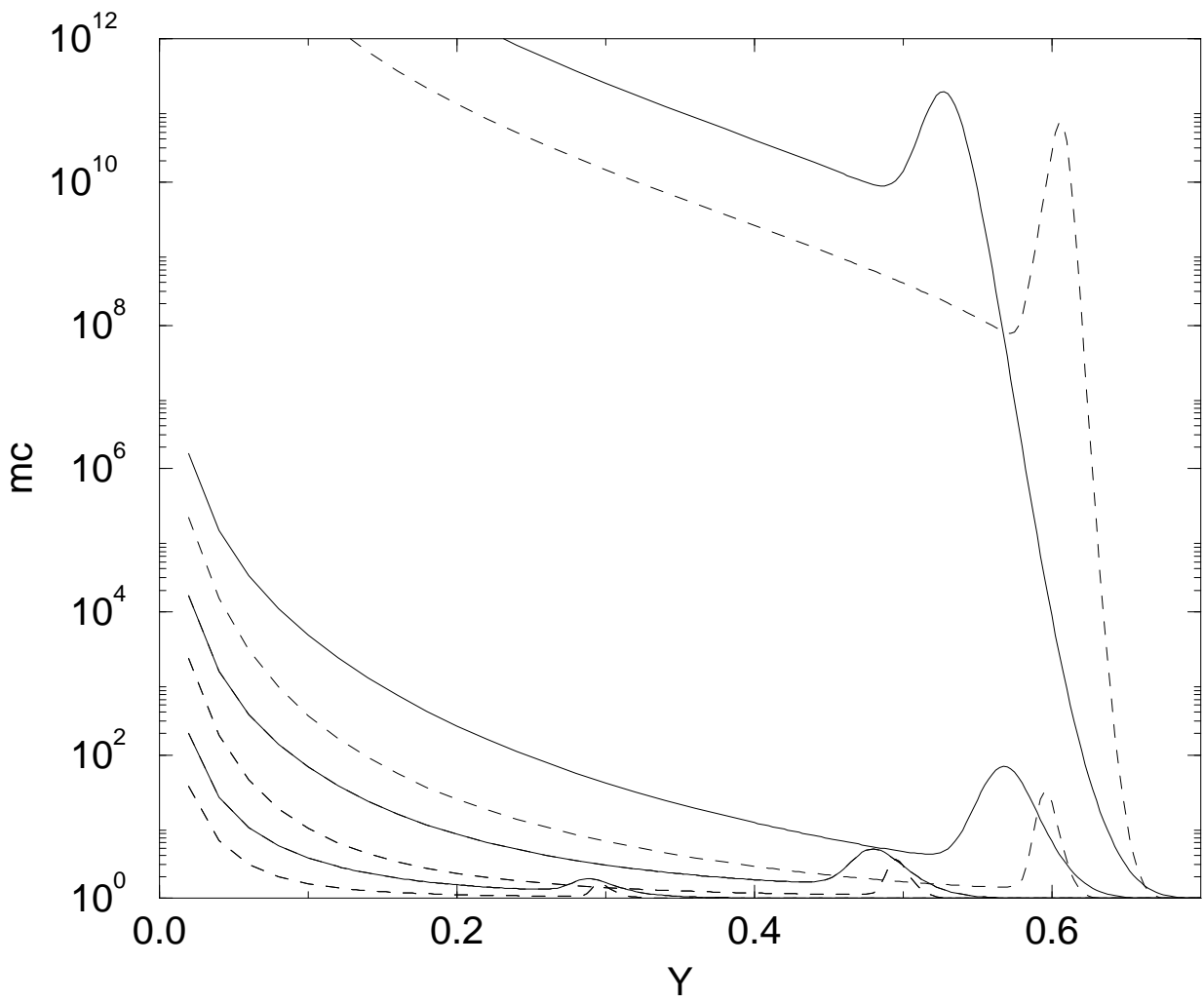
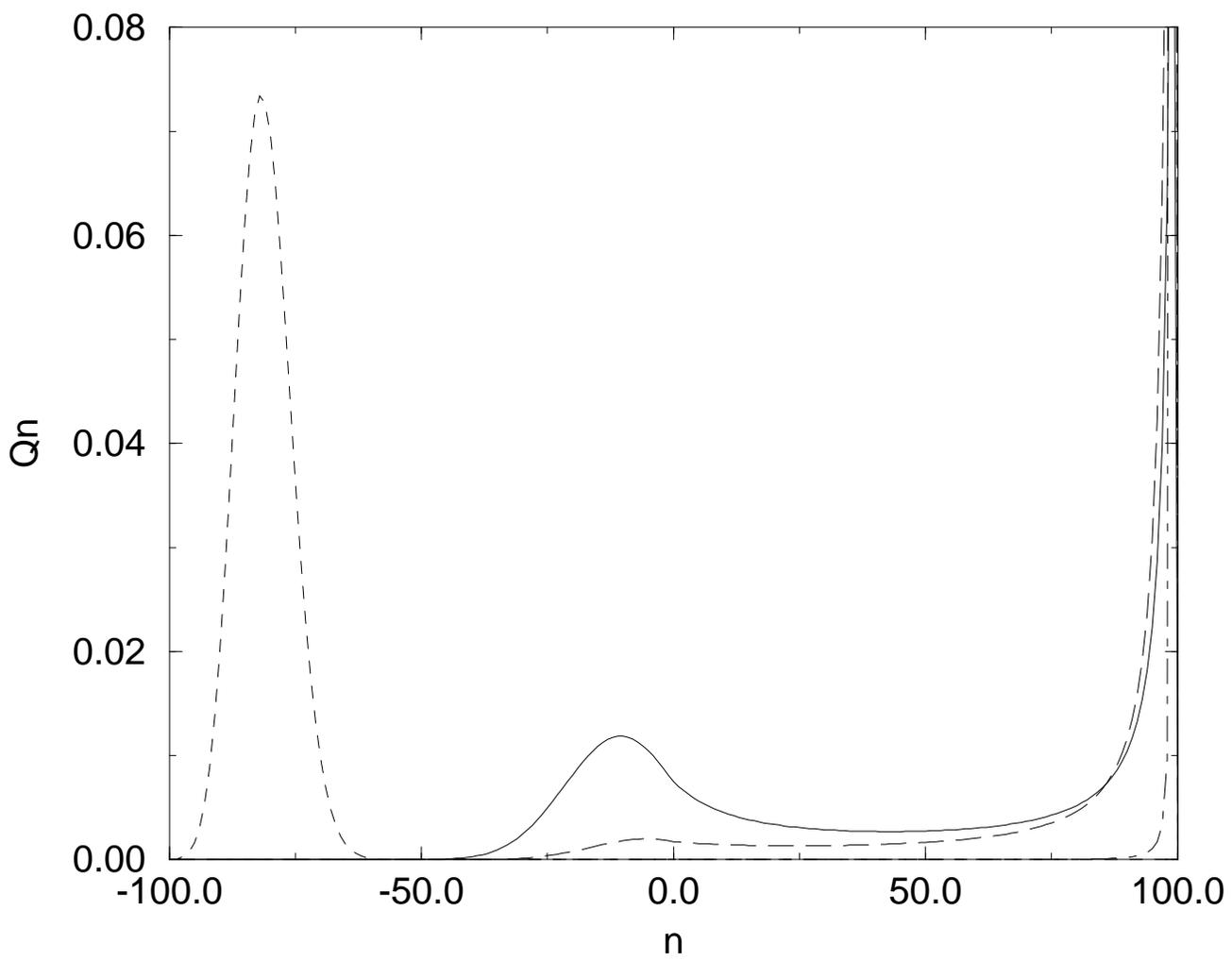
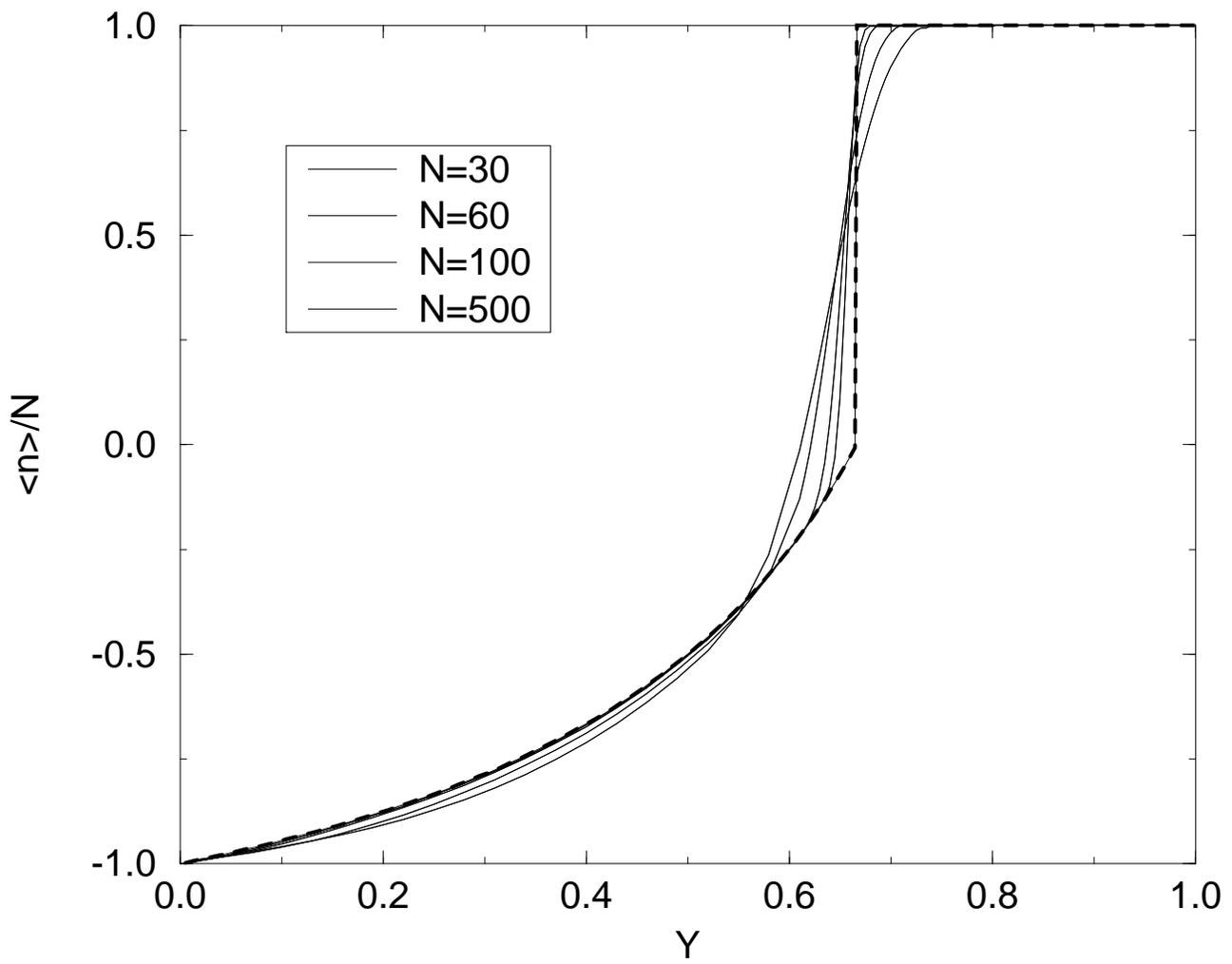
\end{document}